# Monolithic Thulium Fiber Laser with 567 W output power at 1970 nm

Till Walbaum,[1,*] Matthias Heinzig,[1] Thomas Schreiber,[1] Ramona Eberhardt,[1] Andreas Tünnermann[1,2,3]

[1]*Fraunhofer Institute for Applied Optics and Precision Engineering, Albert-Einstein-Str. 7, 07745 Jena, Germany*
[2]*Institute of Applied Physics, Friedrich-Schiller-University Jena, Albert-Einstein-Str. 15, 07745 Jena, Germany*
[3]*Abbe Center of Photonics, Friedrich Schiller University Jena, Max-Wien-Platz 1, 07743 Jena, Germany*
*\*Corresponding author: till.walbaum@iof.fraunhofer.de*

**We report on a monolithic thulium fiber laser with 567 W output power at 1970 nm, which is the highest power reported so far directly from a thulium oscillator. This is achieved by optimization of the splice parameters for the active fiber (minimizing signal light in the fiber cladding) and direct water cooling. Dual transverse mode operation is visible from the optical spectrum and can also be deduced from the measured beam quality of $M^2 = 2.6$.**

Operating in the eye-safe region, Tm-doped fiber lasers have attracted increasing interest during the last years. By spectrally covering or avoiding absorption bands, medical applications, material processing or long-range atmospheric transmission can be addressed using thulium as an active medium. Using sufficiently high concentration of thulium ions, effective diode pumping at 793 nm is possible by exploiting the cross relaxation process [1]. This way, more than 1 kW of output power at 2045 nm has been presented from a MOPA system [2], significantly exceeding the power that has been shown from other eye-safe systems [3]. In comparison to multi-stage amplifier systems, high power oscillators reduce cost and complexity at the expense of some precision in the control over temporal and spectral output parameters. Around 300 W at 2050 nm have been presented from an oscillator including a free-space section [4]. For the spectral region below 2000 nm the output powers shown so far are lower because the maximum gain of the necessary large mode area (LMA) fibers is usually shifted towards longer wavelengths. Consequently, the highest powers below 2 micron wavelength presented so far directly from diode pumped thulium oscillators are 170 W at 1950 nm [5] and 300 W at 1908 nm [6]. Nonetheless, the wavelength range between 1.9 µm and 2 µm is interesting not only for medical applications but also as pump sources for thulium- or holmium-doped fibers [5, 7]. The reduced quantum defect in such so-called tandem pumped systems allows distributing a significant part of the heat load of a high power system from the actual amplifying fiber to the different pump lasers.

In this paper, we present a monolithic thulium fiber oscillator with 567 W output power at 1970 nm wavelength, which is the highest power presented so far from a few-mode thulium-doped fiber laser (TDFL). The operating wavelength is far from the emission maximum of the active fiber (which is around 2030 nm). The scalability to high output power is achieved by carefully optimizing the splices between active and passive fibers (see also [8]) and applying direct water cooling.

The setup of the laser system is depicted in Fig. 1. The monolithic cavity consisted of an active fiber spliced between a highly reflective (HR) fiber Bragg grating (FBG) and one with low reflectivity (LR). Both FBGs were commercially available components, written into passive (non-photosensitive) fiber that matched the active fiber in terms of mode field diameter. The HR-FBG had a reflectivity of 99.6 % with a bandwidth of 3 nm (at 95%), while the LR-FBG had 11 % reflectivity and an FWHM of 1.5 nm. Both FBGs were thermally contacted to a water-cooled heat sink. The active fiber between the FBGs was a highly thulium-doped LMA fiber with 25 µm core and 400 µm cladding diameter (Nufern LMA-TDF-25P/400-HE). A (measured) core NA of 0.1 was enabled by an additional doped pedestal with increased refractive index and 40 µm diameter around the core. The active fiber had a length of ~7 m and was spooled on a water-cooled metal cylinder, which was fully embedded in a water-filled basin. It should be noted that no dedicated effort (heat conductive foil, grooves in the cylinder…) was made to optimize thermal contact between fiber and cylinder to allow for easier exchange of the fiber if necessary. Nonetheless, we tested the free-space pumped setup presented in [8] (with the same fiber cooling) up to 75 W of pump and 26.5 W output power without embedding in the water basin, in which case we were limited by the splice temperature of ~90 °C.

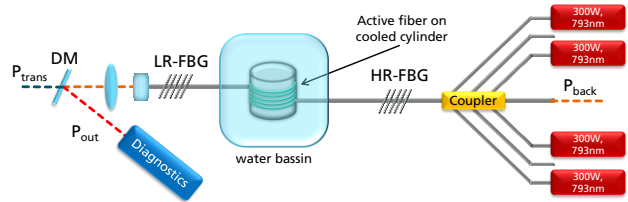

Fig. 1: Experimental setup of the monolithic fiber laser. HR-FBG and LR-FBG: highly reflective and low reflection fiber Bragg grating, respectively; DM: dichroic mirror; coupler: (6+1) x 1 pump fiber coupler with signal feedthrough; Diagnostics: powermeter, optical spectrum analyzer, focus monitor (beam quality measurement).

The splices between active and passive fiber were thus ensured to also be embedded in water and were carefully optimized using refractive index tomography to minimize leaking signal light. For this, the refractive index was measured close to the splice in steps as low as 20 µm along the fiber and across the splice position using a commercially available device based on transverse interferometry (Interfiber Analysis IFA-100, see also [8]). The respective refractive index profiles at the splice position for standard parameters (taken from a splice program for passive fibers) and optimized values (reduced arc duration) are presented in Fig. 2, where a) shows the refractive index profiles far from the splice (2 mm distance) and close to the splice (30 µm distance) for both splice parameters. As can be seen, dopant diffusion leads to significant broadening of the actively doped fiber core if the splice duration is too long, up to a point where the core can hardly be distinguished from the pedestal (Fig. 2a, red curve). It should be noted that camera images from the splicer did not reveal this effect. A more detailed refractive index profile of both splices is depicted in Fig. 2b and Fig. 2d, showing the evolution of the profile towards the splice. Only the pedestal regions are shown there, and a constant offset has been added to refractive index of the passive fiber (right side) for better visibility. Diffusion effects can clearly be seen here. From these measurements, we ran beam propagation simulations starting with the fundamental mode of the active fiber. The results are plotted in Fig. 2c and Fig. 2e for original and optimized splice parameters, respectively. Apart from the obvious excitation of higher order modes for the original splice parameters, the total splice loss calculated from these simulations also amounted for 22.7 % with the original parameters and only 0.5 % with the optimized parameters. It should be noted that the diffusion effects were not visible on the camera image of the splicer. Furthermore, excellent cleaning and relative cleave angles below 1° were also ensured for high power splice connections, and splices were performed using manual alignment.

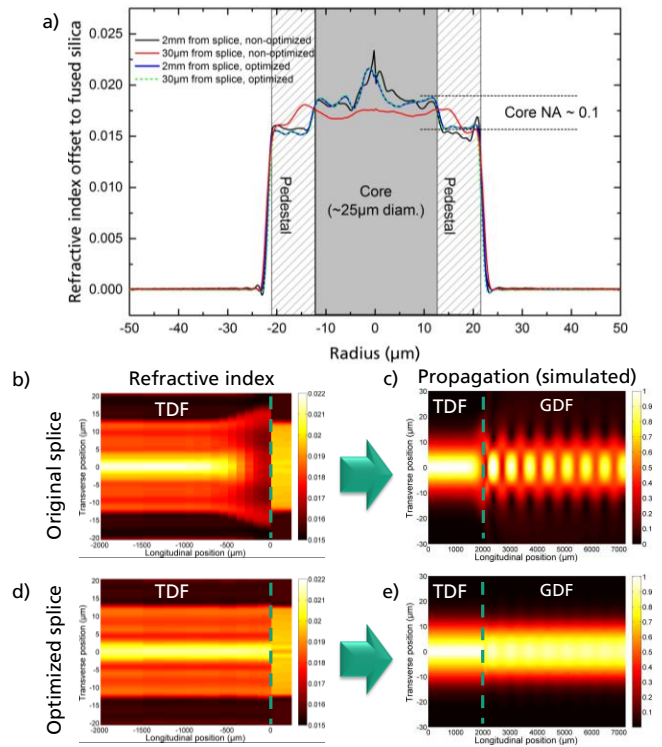

Fig. 2: a) Refractive index profile of the thulium doped fiber in different distances from the splice for original (black and red) and optimized (blue and dashed green) parameters; b) and d) Refractive index evolution (color-coded) along the fiber in the splice region. Active fiber to the left of the splice (green dashed line), passive fiber to the right; c) and e) Beam propagation through the measured refractive index profiles, starting with the fundamental mode of the active fiber on the left side.

Pump power was provided by four 300 W diode lasers (DILAS Compact Evolution) running at 793 nm that were combined via a (6+1) x 1 fiber coupler with signal feedthrough (LightComm) and injected into the cavity through the HR-FBG. At the signal port of the coupler, the backwards propagating power $P_{back}$ could be measured, the fiber end was angle cleaved to reduce reflections. The principal output power was measured behind the LR-FBG, where an AR-coated silica endcap was used to avoid back reflection and protect the fiber facet. A dichroic mirror was used to separate transmitted pump light and signal light, so that both could be measured separately ($P_{trans}$ and $P_{out}$, respectively). For the signal light, an optical spectrum analyzer was employed as well as a focus monitor to determine the beam quality (explained below). We used an additional photo diode to monitor the temporal stability of the signal and scan it for signs of modal instabilities, since no real time beam profiling was available.

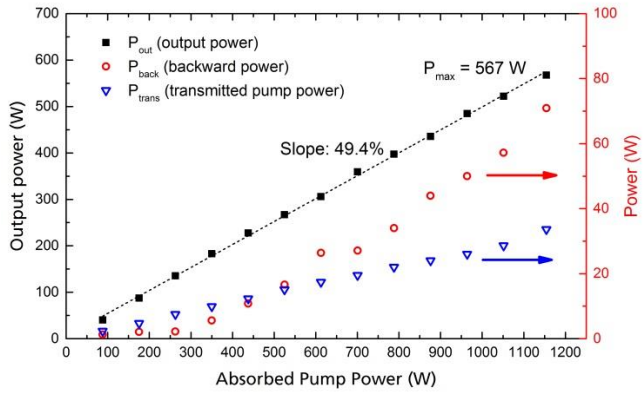

Fig. 3: Forward output power (black, left y-axis) as well as transmitted pump and backwards propagating power (blue and red, respectively, right y-axis) in dependence of the absorbed pump power.

For the high power experiments, the water in the basin was cooled to 10 °C and the cooling water of the heat sink was set to 8 °C. The output characteristics were measured while increasing the pump power, and resulting power data is illustrated in Fig. 3. It can be seen that the output power (forward only) at the signal wavelength increased linearly up to 567 W, which is the highest output power presented so far directly from a thulium-based fiber oscillator below 2 µm and was limited only by available pump power. The system was run at this power for about 20 minutes. The slope of 49.4 % with respect to absorbed pump power was slightly lower than previously reported [5, 6], owing to the output coupling ratio and the fact that the operating wavelength was far from the emission maximum of the active fiber (which was observed to be around 2030 nm). It indicates a quantum efficiency of $\eta = 1.2$. The transmitted pump power reached 33.6 W, which corresponds to 2.8 % of the launched pump power (approximately 2.2 dB / m pump absorption) and indicates that a shorter active fiber could also be used without loss of performance. A maximum of 70.9 W was transmitted through the HR-FBG and measured as backwards propagating power. Due to the high reflectivity of the HR-FBG at the design wavelength, it is to be assumed that spectral broadening leads to higher transmission. Unfortunately, spectral characterization of the backwards propagating light was not possible with the setup.

The signal spectrum is shown in Fig. 4 for different output power levels. As is visible from the figure, the peak emission wavelength shifted from around 1966 nm to over 1970 nm with increasing power. This was due the heat-up of the FBGs, which causes their reflectivity peaks to shift towards longer wavelengths. In fact, in spite of the heat sinks used for the FBGs, the HR-FBG was observed to have 47 °C and the LR-FBG even heated up to 52 °C at the highest power. The bandwidth at 3 dB was 180 pm at the peak wavelength, in spite of the much higher bandwidths of the FBGs. It can also be seen from the figure that, at every power level, the spectrum features at least two pronounced peaks. By measuring the refractive index profile and calculating (by FEM simulation) the transverse modes of the passive fiber, we could identify that these peaks correspond to the fundamental mode and the first higher order mode. Due to the different effective refractive indices for the transverse modes, the peak reflectivity wavelength of the FBGs is different for each of them. Note that no mode conversion is necessary for the HOM to appear here. In fact, mode conversion would result in a third spectral peak between those associated to fundamental mode and first HOM, which was not observed. Tighter fiber bending (currently 10 cm diameter), or a smaller signal core will be used to suppress the HOM in future experiments, this is subject of current investigations. Apart from the two main peaks, two additional peaks become visible at 267 W output power (red curve in the figure, peaks at 1964.6 nm and 1970 nm). Their symmetry suggests that these could be attributed to non-degenerate intermodal four-wave-mixing of the two primary wavelengths. Since they are less pronounced at higher power, it is possible that phase-matching deteriorates due to thermal effects in the fiber.

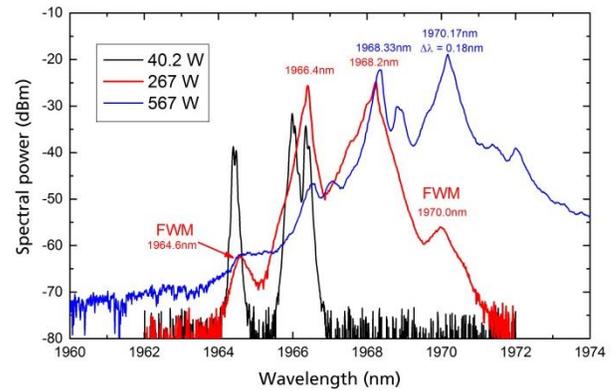

Fig. 4: Optical spectrum of the signal light (forwards propagating) for different output power levels (black: 40.2 W; red: 267 W; blue: 567 W). FWM demarks four-wave-mixing peaks.

The significant higher order mode content also showed in the beam quality of the signal. To observe the beam profile, a focus monitor (PRIMES FocusMonitor FM120) was used. The measurement principle is based on a quickly (7500 rpm, 125 Hz) rotating needle which moves an attached pinhole (~20 µm diameter) across the beam transversely. By scanning the rotating needle's position, a beam profile can be calculated from the measured power data. This method has the advantage that Watt-level powers can be directly observed, as opposed to the milli- or microwatt levels that can be monitored on cameras for this wavelength. However, since it is basically a scanning measurement, no real time imaging is possible and rapid beam profile fluctuations cannot be reliably identified. The beam quality at maximum output power has been measured with this device, and the results are shown in Fig. 5. A beam quality factor of $M^2=2.6$ was extracted from the measured beam profiles. According to the literature [9], such an $M^2$ would be expected for a HOM content of about 70-80 %; however, the recorded spectral data (Fig. 4) suggests that the HOM should not be more than 50 %, which would correspond to an $M^2$ of 1.3 - 1.8. We assume that beam fluctuations due to thermal effects in the air and thermal lensing are responsible for this, which could not be identified due the lack of real-time capability of the beam profiler, but have been reported previously for Tm-based systems [10]. This is also supported by the fact that the longitudinal focus position in the

M$^2$-measurement changed with increasing power. Apart from the propagation through the air, the AR-coated fiber end cap is the most probable source for thermal lensing in the setup. We observed that it reached a temperature of ~70 °C at the highest output power, while the other optics remained relatively cool.

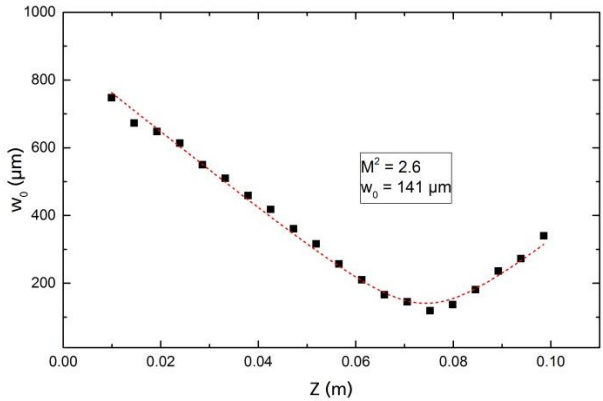

Fig. 5: Beam quality measurement of the thulium oscillator at 567 W output power.

In conclusion, we have presented an all-fiber oscillator based on thulium-doped active fiber with a record output power of 567 W at 1970 nm. This was achieved by combining optimized fiber-to-fiber connections and direct water cooling. Spectral data and beam quality measurement revealed dual-transverse mode operation, which should not be an issue for tandem pumping applications, but will be eliminated anyway as the next step.

**Funding.** European Research Council (ERC) (670557 "MIMAS").

## Full References